\documentclass{article}
\usepackage[preprint]{spconf,amsmath,graphicx}
\usepackage{amssymb}
\usepackage{ dsfont }
\usepackage{booktabs}
\usepackage{hyperref}
\usepackage{subcaption}
\usepackage{cite}
\usepackage{comment}
\usepackage{tikz}

\newcommand\copyrighttext{%
  \tiny Copyright 2023 IEEE. Published in the 2022 IEEE Spoken Language Technology Workshop (SLT) (SLT 2022), scheduled for 19-22 January 2023 in Doha, Qatar. Personal use of this material is permitted. However, permission to reprint/republish this material for advertising or promotional purposes or for creating new collective works for resale or redistribution to servers or lists, or to reuse any copyrighted component of this work in other works, must be obtained from the IEEE. Contact: Manager, Copyrights and Permissions / IEEE Service Center / 445 Hoes Lane / P.O. Box 1331 / Piscataway, NJ 08855-1331, USA. Telephone: + Intl. 908-562-3966.}
\newcommand\copyrighttextinbottomleft{%
\begin{tikzpicture}[remember picture,overlay]
\node[anchor=south,yshift=20pt] at (current page.south) {\parbox{\dimexpr\textwidth-\fboxsep-\fboxrule\relax}{\copyrighttext}};
\end{tikzpicture}%
}

\newcommand{\norm}[1]{\left\lVert#1\right\rVert}
\title{Improving generalizability of distilled self-supervised speech processing models under distorted settings}
\name{%
\begin{tabular}{@{}c@{}}
Kuan-Po Huang$^{1\star}$ \qquad 
Yu-Kuan Fu$^{2\star}$ \qquad 
Tsu-Yuan Hsu$^{3\star}$ \qquad 
\thanks{$^{\star}$Equal contribution.}
Fabian Ritter Gutierrez$^{4\dagger}$\thanks{$^\dagger$Work done as a research engineer at National University of Singapore.}\\
Fan-Lin Wang$^{5}$ \qquad 
Liang-Hsuan Tseng$^{6}$ \qquad 
Yu Zhang$^{7}$ \qquad 
Hung-yi Lee$^{8}$
\end{tabular}}
\address{
$^{123568}$National Taiwan University \qquad
$^{1}$ASUS Intelligent Cloud Services\\
$^{4}$Nanyang Technological University \qquad
$^{7}$Google Brain \\
$^{123568}$\{f09922005, b07202024, b08201047, b07502100, b07502072, hungyilee\}@ntu.edu.tw\\
$^{4}$s220064@e.ntu.edu.sg,
$^7$ngyuzh@google.com
}
\begin{document}
\ninept
\maketitle
\copyrighttextinbottomleft
\begin{abstract}
Self-supervised learned (SSL) speech pre-trained models perform well across various speech processing tasks. Distilled versions of SSL models have been developed to match the needs of on-device speech applications. Though having similar performance as original SSL models, distilled counterparts suffer from performance degradation even more than their original versions in distorted environments. This paper proposes to apply Cross-Distortion Mapping and Domain Adversarial Training to SSL models during knowledge distillation to alleviate the performance gap caused by the domain mismatch problem. Results show consistent performance improvements under both in- and out-of-domain distorted setups for different downstream tasks while keeping efficient model size. 

\end{abstract}
\section{Introduction}
\label{sec:intro}
Self-supervised learned (SSL) speech pre-trained models~\cite{mohamed2022self}, unlike traditional models, eliminated the need for labeled speech data during training. Large SSL models, such as Wav2vec2.0 \cite{baevski2020wav2vec} and HuBERT \cite{hsu2021hubert}, are known for generating speech representations that perform well in different downstream speech processing tasks on the Speech processing Universal PERformance Benchmark \cite{yang2021superb} (SUPERB). Though having great success in performance for numerous speech processing tasks, the sizes of the existing SSL models narrow the general usage of them, especially on on-device applications where memory and computation resources are limited. To overcome this disadvantage of SSL models, DistilHuBERT \cite{chang2022distilhubert}, a model distilled from HuBERT with knowledge distillation, was developed to have a much smaller model size while preserving the performance of downstream speech processing tasks at a certain level.

Another severe problem is the generalizability of SSL models. Domain shifts caused by mismatches between training data and testing data usually occur in real-world scenarios. A common factor that causes domain shifts is speech distortions. 
In this paper, we focus on the setting that the training data of downstream tasks contain clean speech while the testing data has distortions. 
For many downstream speech processing tasks such as Intent Classification (\textbf{IC}), Keyword Spotting (\textbf{KS}), Emotion Recognition (\textbf{ER}), and Automatic Speech Recognition (\textbf{ASR}), since the cost of collecting labelled training data is expensive, the training dataset can not be diverse, and usually only has clean speech. 
However, there may be background noises during the testing phase in real-world applications, making SSL models vulnerable in performance as studied in \cite{huang2022improving}.



In this paper, we find that although distilled SSL models have comparable performance with their teachers, distilled SSL models suffer from performance degradation even more than their teachers in distorted environments.
To overcome the problem that distilled models are especially vulnerable to distorted speech, we propose to apply Cross-Distortion Mapping (CDM) during knowledge distillation to improve the generalizability of DistilHuBERT. 
The process of Cross-Distortion Mapping refers to a teacher-student learning framework with the teacher and student model having different distorted inputs. 
The results show that CDM improves the testing performance on downstream speech processing tasks under the setting with speech distortions, even when the distortion types are unseen during training.  
To further improve the robustness of the teacher model, we performed domain-adaptive pre-training on the teacher model by utilizing distorted pre-training data so that the student model would be able to have a more robust target to learn with. 
We also applied Domain Adversarial Training~\cite{ganin2016domain} (DAT) in the hope of generating more domain-invariant speech representations, and found out that DAT benefits the generalization of models in some cases\footnote{Code will be released at \href{https://github.com/nobel861017/distort-robust-distilSSL}{https://github.com/nobel861017/distort-robust-distilSSL}.}. 

\section{Related work}

There are several studies enhancing the robustness of SSL models. 
It has been found that pre-training some more steps with unlabeled target domain data \cite{hsu2021robust} on Wav2vec2.0 mitigates the problem of domain shifts. Therefore an intuitive method to enhance the robustness of SSL models is to augment pre-trained data by adding distortions.

In this paper, we found that the proposed CDM method further improved the distilled SSL model learned from the domain-adapted teacher model pre-trained with distorted speech.

Besides domain-adaptive pre-training, DAT is applied to improve the generalizability of SSL models. 
Augmentation adversarial training \cite{huh2020augmentation} combines the concept of augmentation and DAT to generate representations invariant to augmentations. 
Some other studies also apply DAT to adapt models to different kinds of data and are also proved to benefit unseen domains, such as different accented speech \cite{ hu2021redat, das2021best} or distorted speech \cite{liao2018noise, huang2022improving}.
Based on our experimental results, DAT sometimes further improved the results when combined with our proposed CDM method.

Several studies have successfully improved the robustness of Wav2vec2.0 models. 
Having noisy waveforms as input, performing clean speech reconstruction with a reconstruction module along with pre-training \cite{wang2022improving} also improves noise robustness. 
Having both clean and noisy speech as input, a denoising approach\cite{zhu2022noise} conducted by constructing clean quantized vectors serving as the target for the noisy representations successfully improved performance on noisy testing sets while preserving performance on the original clean testing set. 
The idea of \cite{zhu2022noise} is similar to setup1 of CDM (will be elaborated in Section~\ref{ssec:cdm}), where the student model learns to generate clean representations of the teacher given distorted speech inputs.
However, we find that setup2 of CDM, where both student and teacher models are given distorted speech, is more effective in gaining robustness than setup1.

CDM is also similar to another series of work, Bootstrap Your Own Latent (BYOL) \cite{grill2020bootstrap}, which is proposed for image processing but also shown to be useful for audio~\cite{niizumi2021byol} and speech~\cite{data2vec, elbanna2022byol} applications.
BYOL performs augmentation to data and trains an online-target framework from scratch.
The online-target framework consists of an online network and target network with similar model architectures. Both networks receive the same input but with different augmentations and minimize the distance of output representations between the two. 
As BYOL, CDM includes the augmentation method of BYOL and the concept of denoising by viewing different distorted speech as different domains and minimizing the distance of speech belonging to different domains. 
Different from BYOL, we utilized a teacher-student model compression framework where a small student model is trained to mimic a large teacher model.

\section{Methods}

\begin{figure*}
    \centering
    \includegraphics[width=17cm]{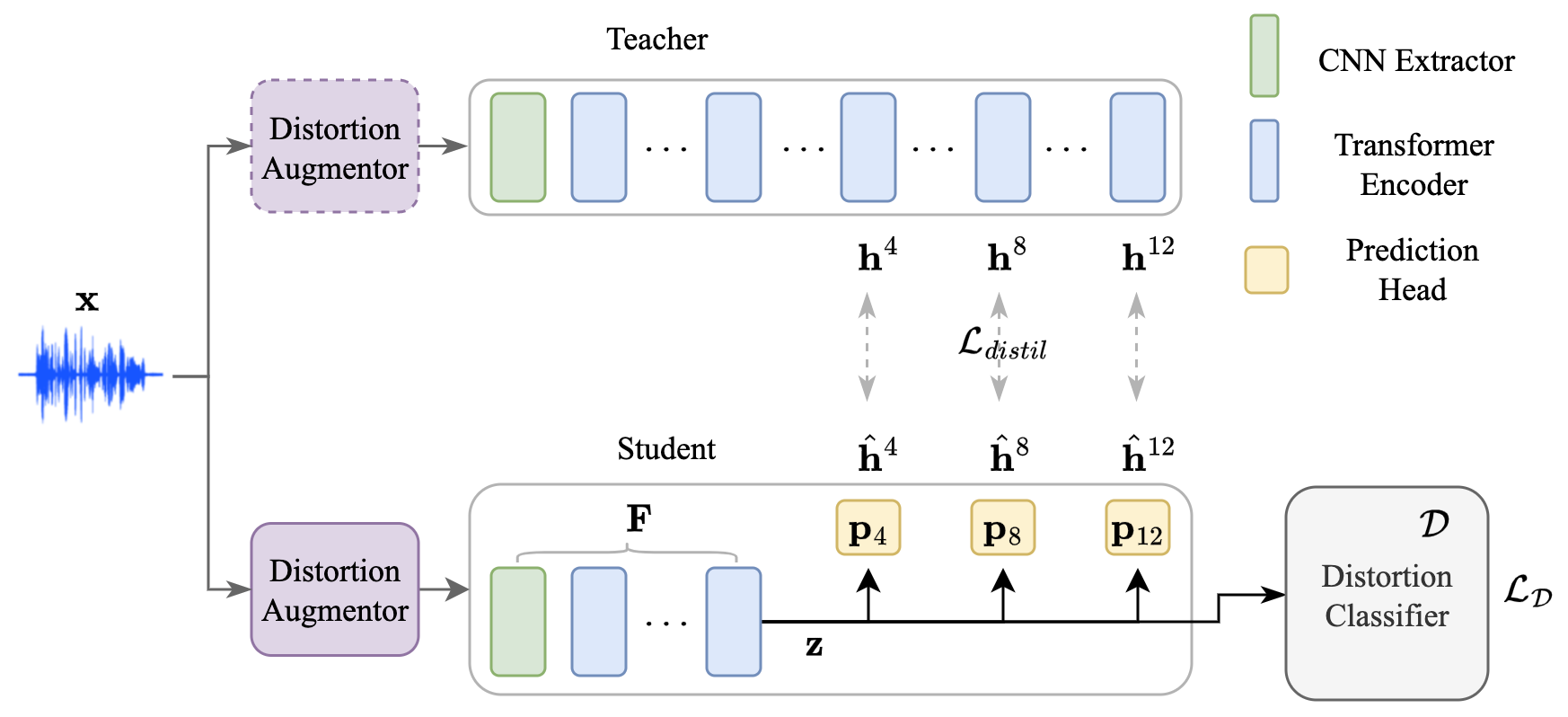}
    \caption{Illustration of DistilHuBERT with a distortion classifier. The distortion augmentor follows the procedure mentioned in Section \ref{ssec:dataprep}. For setup1, there is no distortion augmentor for the teacher model. For setup2, both the teacher and student model have a distortion augmentor in front.}
    \label{fig:DistilHuBERT_DAT}
\end{figure*}
\label{sec:methods}

\subsection{Knowledge distillation}
\label{ssec:kd}
Throughout this work, DistilHuBERT~\cite{chang2022distilhubert} is adopted to meet our needs for reducing model size. DistilHuBERT is trained with a teacher-student learning framework with knowledge distillation. As shown in Fig.~\ref{fig:DistilHuBERT_DAT}, the student network consists of a subnet $\mathbf{F}$ followed by some prediction heads $\mathbf{p}$, where $\mathbf{F}$ is constructed by reducing the number of transformer encoder layers of the HuBERT teacher model. Given an input speech utterance $\mathbf{x} \in \mathds{R}^{T}$, where $T$ is the number of timesteps of $\mathbf{x}$, a predicted hidden representation sequence $\hat{\mathbf{h}}^i$ is output by the student model as shown in Eq.~(\ref{eq:stu_forward}). 
\begin{equation}
\label{eq:stu_forward}
    \begin{aligned}
        \mathbf{z} &= \mathbf{F}(\mathbf{x})\\ \hat{\mathbf{h}}^i &= \mathbf{p}_i (\mathbf{z})
    \end{aligned}
\end{equation}
In Eq.~(\ref{eq:stu_forward}), $\mathbf{z}$ is the last hidden representation of the transformer layers in the student model, and serves as the input of the prediction heads $\mathbf{p}$. Prediction head $\mathbf{p}_i$ predicts the $i^{th}$ hidden layer representation $\mathbf{h}^{i}$ of the teacher model. The overall objective of DistilHuBERT consists of a $L_1$ loss term and a cosine similarity loss term as shown in Eq. (\ref{eq:distil_loss}), 
\begin{equation}
\label{eq:distil_loss}
\begin{aligned}
    & \mathcal{L}_{distil} = \mathcal{L}_{L_1} + \mathcal{L}_{cos} \\ =& \sum_{i\in\{4,8,12\}} \sum_{t=1}^T \biggr[\frac{1}{D}\norm{\mathbf{h}_t^{i} - \hat{\mathbf{h}}_t^{i}}_1 - \gamma \log \sigma (\cos(\mathbf{h}_t^{i}, \hat{\mathbf{h}}_t^{i}))\biggr]
\end{aligned}
\end{equation}
where $\sigma$ is the sigmoid function and $\cos(\cdot, \cdot)$ is the cosine similarity function, $D$ is the feature dimension of the representations, and $\gamma$ is a constant to scale the value of the cosine similarity loss. Though both loss terms may have similar goals, the original paper \cite{chang2022distilhubert} reports that considering both terms results in better performance.

\subsection{Model generalization}
\label{ssec:mg}

\subsubsection{SSL model domain-adaptive pre-training}
\label{ssec:ssl_cont.}
Pre-training SSL models with target domain data is an intuitive way to adapt SSL models to another domain \cite{hsu2021robust,huang2022improving,wang2022improving,gururangan-etal-2020-dont}. This method is sometimes referred to as domain-adaptive pre-training or continual training, depending on its training setting and data configuration\footnote{Continual training is performed under a life-long learning scheme, and the data of previous tasks are usually unavailable, which is not the case in our work.}. For some experiments in our work, we trained the pre-trained teacher model with distorted speech data for additional steps to enhance robustness. The distorted data is generated by adding distortions to the original pre-training data of the corresponding SSL model. We hope that in this way, the teacher model will be able to output more robust representations for the student model to learn with. To avoid any possible misunderstanding, we refer to this method as domain-adaptive pre-training in the further sections. Note that this method differs from Domain Adversarial Training mentioned in Section \ref{ssec:dat}, though both are domain-adaptive methods.

\subsubsection{Cross-Distortion Mapping (CDM)}
\label{ssec:cdm}
Cross-Distortion Mapping refers to the augmentation procedure for the teacher-student framework. The teacher and student model receive the same speech utterance augmented with different distortions as shown in Fig.~\ref{fig:DistilHuBERT_DAT}. We investigate two setups for augmentation.
\begin{itemize}
    \item setup1: The first setup is to input speech utterances without distortions to the teacher model, while the student model takes distorted speech as input. This setup has the denoising concept since the distorted speech representations have clean speech representations as targets, and there are previous works \cite{wang2022wav2vec, zhu2022noise, zhu2022joint} showing that this setup is beneficial for training models robust to noise.
    \item setup2: The second setup is to apply different distortions to speech similar to \cite{grill2020bootstrap, niizumi2021byol, chen2021exploring}, resulting in the teacher and student model observing same speech utterances but with different distortions. 
\end{itemize}



\subsubsection{Domain Adversarial Training (DAT)}
\label{ssec:dat}
 DAT is performed by utilizing a distortion classifier $\mathcal{D}$ that takes the last hidden states $\mathbf{z}$ of the student model as input. The distortion classifier aims at identifying the distortion types of the distorted speech input by optimizing a multi-label cross-entropy loss (each speech utterance may be distorted with multiple distortions).

During the training process, the distortion classifier and the student model are trained in turn. First, the parameters of the distortion classifier $\theta_{d}$ is updated with gradient descent shown below, 
\begin{equation}
    \theta_{d}\xleftarrow{} \theta_{d} - \alpha \frac{\partial \mathcal{L}_{D}}{\partial \theta_{d}}
\end{equation}
where $\theta_{\mathcal{D}}$ is the parameters of the distortion classifier, $\mathcal{L}_\mathcal{D}$ is the cross-entropy loss, and $\alpha$ is the learning rate. 

After training the distortion classifier, the parameters $\theta_{s}$ of the student model are updated through the process in the following, \begin{equation}
    \theta_{s}\xleftarrow{} \theta_{s} - \beta \frac{\partial(\mathcal{L}_{distil} - \lambda \mathcal{L}_{D})}{\partial \theta_{s}}
\end{equation}
where $\theta_{s}$ is the parameters of the student model, $\beta$ is the learning rate, and $\lambda$ is a constant controlling the scale of $\mathcal{L}_\mathcal{D}$.

\section{Experimental setup}
\begin{table*}[t]
\setlength\tabcolsep{2.3pt}
\renewcommand{\arraystretch}{1.2}
\centering
\begin{tabular}{llcc|cccc|cccc|cccc}
    \toprule
  &  & \multicolumn{1}{l}{} && \multicolumn{4}{|c}{\textbf{KS} (Acc$\%\uparrow$)}  & \multicolumn{4}{|c}{\textbf{IC} (Acc$\%\uparrow$)} & \multicolumn{4}{|c}{\textbf{ER} (Acc$\% \uparrow$)}  \\
  &  & da. & \multicolumn{1}{c}{para.} & \multicolumn{1}{|c}{clean} & \multicolumn{1}{c}{2-dist} & \multicolumn{1}{c}{fsd} & \multicolumn{1}{c}{dns} & \multicolumn{1}{|c}{clean} & \multicolumn{1}{c}{2-dist} & \multicolumn{1}{c}{fsd} & \multicolumn{1}{c}{dns} & \multicolumn{1}{|c}{clean} & \multicolumn{1}{c}{2-dist} & \multicolumn{1}{c}{fsd} & \multicolumn{1}{c}{dns} \\
  \midrule
(T1)  & HuBERT \cite{hsu2021hubert}   & X& 95M  & 96.30 & 89.81   & 90.94 & 77.60   & 98.34 & 89.09   & 91.93 & 74.11  & 64.92& 56.72   & 60.05 & 52.08  \\
(T1') & HuBERT   & V& 95M  & 96.53& 94.77   & 94.00   & 82.83  & 98.37& 96.20& 96.78 & 85.00& 65.88& 62.82   & 63.89 & 56.70   \\
(S1)  & DistilHuBERT (Tr2) \cite{chang2022distilhubert}   & X& 23M  & 95.98& 87.57   & 88.70  & 75.07  & 94.99& 70.29   & 72.50  & 48.30   & 63.13& 55.09   & 57.05 & 49.76  \\
(S1') & DistilHuBERT (Tr2)   & V& 23M  & 96.14& 86.86   & 90.56 & 76.47  & 95.65& 77.99   & 81.73 & 57.50   & 64.01& 58.89   & 59.06 & 53.14  \\
\midrule
(S2)  & DistilHuBERT (Tr2) setup1   & X& 23M  & 95.52& 92.92   & 93.44 & 76.66  & 94.17& 89.53   & 89.61 & 72.11  & 63.51& 58.11   & 60.17 & 50.66  \\
(S2') & DistilHuBERT (Tr2) setup1   & V& 23M  & 96.17& 93.61   & 94.09 & 77.44  & 95.57& 86.11   & 89.03 & 71.26  & 63.72& 59.62   & 61.42 & 53.69  \\
\midrule
(S3)  & DistilHuBERT (Tr2) setup2 (same) & X& 23M  & 96.11 & 89.84   & 91.69 & 78.42  & 94.62 & 75.40 & 80.33 & 57.92  & 61.87& 55.72   & 59.41 & 50.27  \\
(S3') & DistilHuBERT (Tr2) setup2 (same) & V& 23M  & 96.33& 92.57   & 93.48 & \textbf{80.04}  & 95.68& 85.16   & 86.84 & 64.46  & \textbf{64.25} & 59.62   & 60.93 & 51.78  \\

(S4)  & DistilHuBERT (Tr2) setup2   & X& 23M  & 96.27& 92.99   & 93.96 & 77.47  & 95.91 & 90.72   & 90.77 & 73.87  & 63.77& 59.89   & 61.62 & 51.25  \\
(S4') & DistilHuBERT (Tr2) setup2   & V& 23M  & \textbf{96.53} & 93.61   & 94.38 & 79.10   & 96.57& \textbf{92.25}   & \textbf{92.67} & \textbf{78.41}  & 63.08& 60.38   & 60.89 & 53.38  \\
\midrule
(S5)  & DistilHuBERT (Tr2) setup1 + DAT & X& 23M  & 95.94& 93.80& 93.83 & 79.36  & 96.02& 90.35   & 91.09 & 74.61  & 63.41& 59.34   & 60.58 & 53.29  \\
(S5') & DistilHuBERT (Tr2) setup1 + DAT & V& 23M  & 95.75& 93.61   & 93.35 & 78.25  & 96.34 & 88.82   & 90.48 & 73.19  & 63.23 & 60.06   & 61.15 & \textbf{53.71}  \\
(S6)  & DistilHuBERT (Tr2) setup2 + DAT & X& 23M  & 96.17& 93.77   & 93.90  & 78.45  & 96.49& 91.35   & 92.09 & 75.51  & 63.44& 59.36   & \textbf{61.82} & 51.01  \\
(S6') & DistilHuBERT (Tr2) setup2 + DAT & V& 23M  & 96.46& \textbf{94.03}   & \textbf{94.55} & 78.90   & \textbf{96.75} & 91.01   & 92.06 & 76.51  & 63.45& \textbf{61.15}   & 61.62 & 53.49  \\
\midrule
(S7)  & DistilHuBERT (Tr1)   & X& 20M  & 94.90 & 86.34   & 87.47 & 71.44  & 92.35& 60.43   & 64.25 & 38.65  & 62.45& 52.65   & 57.19 & 49.23  \\
(S7') & DistilHuBERT (Tr1) setup2  & V& 20M  & 96.46&	92.79&	93.44& 75.33&	94.96&	84.66&	86.29&	64.36& 62.93&	58.56&	59.53&	50.43\\
(S8)  & DistilHuBERT (Tr3)   & X& 34M  & 96.53& 89.39   & 90.85 & 76.50   & 94.70 & 74.00 & 78.12 & 54.15  & 62.94& 55.34   & 56.94 & 51.69\\
(S8')  & DistilHuBERT (Tr3) setup2 & V& 34M & 96.53&	93.90&	94.61&	77.90 & 97.47 & 93.49 & 93.80 & 79.57 & 64.63&	62.88&	63.25&	53.98\\
\bottomrule
\end{tabular}
\caption{Evaluation results for \textbf{KS}, \textbf{IC}, and \textbf{ER} in accuracy (Acc). By default, DistilHuBERT and has two transformer encoder layers (Tr2). Tr1 and Tr3 denote the number of transformer encoder layers (1 and 3) of DistilHuBERT, which are different from the default configuration (Tr2). The second column, da., specifies whether domain-adaptive pre-training is conducted to the teacher model. The third column, para., lists the number of parameters for each model. The terms ``setup1" and ``setup2" refer to the two setups of the CDM method mentioned in Section \ref{ssec:cdm}. The best performance on each test set throughout the twelve DistilHuBERT models in (S1)-(S6') is marked in bold.}
\label{tab:KS_IC_ER}
\end{table*}
\begin{table*}[ht]
\setlength\tabcolsep{2.0pt}
\renewcommand{\arraystretch}{1.15}
\centering
\begin{tabular}{llcc|cc|cc|cc|cc|cc|cc}
    \toprule
  &  & \multicolumn{1}{l}{} && \multicolumn{12}{|c}{\textbf{ASR} (WER$\% \downarrow$)}  \\
  &  & da. & \multicolumn{1}{c}{ para} & \multicolumn{1}{|c}{clean} & \multicolumn{1}{c}{LM} & \multicolumn{1}{|c}{other} & \multicolumn{1}{c}{LM} & \multicolumn{1}{|c}{2-dist} & \multicolumn{1}{c}{LM} & \multicolumn{1}{|c}{fsd} & \multicolumn{1}{c}{LM} & \multicolumn{1}{|c}{dns} & \multicolumn{1}{c}{LM} & \multicolumn{1}{|c}{CHiME3} & \multicolumn{1}{c}{LM} \\
  \midrule
(T1)  & HuBERT \cite{hsu2021hubert}   & X& 95M  & 6.42&	4.79& 15.67&	12.14&	11.28&	8.59&	9.36&	7.20&	22.72&	18.87&	24.74& 20.28  \\
(T1') & HuBERT  & V& 95M  & 6.75&	4.91& 16.01&	12.44&	8.31&	6.17&	7.78&	5.86&	16.31&	12.85&		21.98&	17.51\\
(S1)  & DistilHuBERT (Tr2) \cite{chang2022distilhubert}   & X& 23M  & 13.37&	9.21& 35.65&	27.78&	30.65&	24.32&	24.27&	18.37&	54.46&	47.52&		46.62&	37.68  \\
(S1') & DistilHuBERT (Tr2)   & V& 23M  & \textbf{13.14} &	\textbf{9.08} &	\textbf{33.97} &	\textbf{26.63} & 26.02&	19.97& 21.80&	16.28&	48.29&	41.70&		43.76&	36.38  \\
\midrule
(S2)  & DistilHuBERT (Tr2) setup1   & X& 23M  & 13.96& 9.64& 36.23&	28.67&	18.52&	13.25&	16.69&	11.87&	38.63&	31.39&	38.15&	29.92  \\
(S2') & DistilHuBERT (Tr2) setup1   & V& 23M  & 13.54 &	9.41& 34.37&	27.18&	\textbf{17.58} &	12.71&	16.38&	11.61&	37.41&	30.66&	37.92&	29.74  \\
\midrule
(S3)  & DistilHuBERT (Tr2) setup2 (same) & X& 23M  & 13.77&	9.52& 35.87&	28.23&	23.49&	17.34&	19.83&	14.36&	45.18&	37.69&		41.30&	32.73  \\
(S3') & DistilHuBERT (Tr2) setup2 (same) & V& 23M  & 13.57&	9.34& 34.34&	26.91&	18.71&	13.33&	17.06&	12.16&	38.77&	31.29&		38.25&	29.94  \\

(S4)  & DistilHuBERT (Tr2) setup2   & X& 23M  & 14.33&	9.90& 36.99&	28.95&	18.58&	13.44&	17.18&	12.13&	39.14&	32.11&		39.13&	30.80  \\
(S4') & DistilHuBERT (Tr2) setup2   & V& 23M  & 13.7&	9.66&	\textbf{33.97} &	27.03& 17.89&	12.98&	16.35&	11.65&	37.73&	31.06&		38.02&	30.18 \\
\midrule
(S5)  & DistilHuBERT (Tr2) setup1 + DAT & X& 23M  & 13.64&	9.40& 34.53&	27.27&	17.86&	12.66&	16.45&	11.72&	37.56&	30.53&		37.63&	30.17  \\
(S5') & DistilHuBERT (Tr2) setup1 + DAT & V& 23M  & 13.72&	9.57& 34.08&	26.82&	17.74&	12.70&	16.30&	11.61&	37.30&	30.43&		\textbf{37.21} &	\textbf{29.31}  \\
(S6)  & DistilHuBERT (Tr2) setup2 + DAT & X& 23M  & 14.46&	10.02& 37.37&	29.41&	18.96&	13.69&	17.52&	12.30&	39.03&	31.77&		39.26&	31.01  \\
(S6') & DistilHuBERT (Tr2) setup2 + DAT & V& 23M  & 13.58&	9.45& 34.07&	26.97&	17.61&	\textbf{12.63} & \textbf{16.16} &	\textbf{11.60} &	\textbf{37.11} &	\textbf{30.33} &	37.57&	29.71  \\
\midrule
(S7)  & DistilHuBERT (Tr1)   & X& 20M  &  14.70&	10.03&	39.73&	31.73& 35.97&	29.33&	28.12&	21.41&	62.91&	57.22&		52.48&	43.44 \\
(S7') & DistilHuBERT (Tr1) setup2  & V& 20M  & 15.20&	10.41& 38.37&	30.61&	20.37&	14.64&	18.69&	13.40&	44.26&	37.07&	42.80&	34.33\\
(S8)  & DistilHuBERT (Tr3)   & X& 34M  & 12.61&	8.73& 32.50&	25.24&	28.05&	21.95&	22.05&	16.54&	51.28&	45.07&	44.06&	36.37 \\
(S8')  & DistilHuBERT (Tr3) setup2   & V& 34M & 12.12&	8.48& 30.92&	23.95&	15.37&	10.92&	14.20&	10.22&	32.08&	25.71&	33.94&	26.63 \\
\bottomrule
\end{tabular}
\caption{Evaluation results for \textbf{ASR} in word error rate (WER). Results of the test-clean set of LibriSpeech is abbreviated as clean, and the results of the test-other set of LibriSpeech is abbreviated as other. LM represents the results after language model rescoring. Notations are same as Table \ref{tab:KS_IC_ER}.}
\label{tab:ASR}
\end{table*}

\subsection{Data preparation}
\label{ssec:dataprep}
The corpus used for knowledge distillation is LibriSpeech \cite{panayotov2015librispeech} 960-hour, which is same as the pre-training data of HuBERT-base in \cite{hsu2021hubert}. In our distorted setting (denoted as 2-dist in Table \ref{tab:KS_IC_ER} and \ref{tab:ASR}), we consider clean speech and speech containing one or two distortions. Distorted speech is generated by applying either one of the additive distortions or one of the non-additive distortions, or both to speech. 

Additive distortions are noises directly added to speech data at a specific speech-noise ratio (SNR) between 10 dB and 20 dB. We adopted additive noise from four widely-known noise datasets, Musan \cite{snyder2015musan}, WHAM! \cite{wichern2019wham}, FSD50k \cite{fonseca2022FSD50K}, and DNS\footnote{We follow the procedure in the original paper \cite{dubey2022icassp} to generate noisy data.} \cite{dubey2022icassp}. Apart from the recorded noise data of the aforementioned datasets, we also took advantage of Gaussian noise, a hand-crafted noise that follows the Gaussian distribution in the time domain. During testing, we evaluated models on four downstream speech processing tasks, \textbf{KS}, \textbf{IC}, \textbf{ER}, and \textbf{ASR}. Besides the original testing set configured by the SUPERB benchmark (denoted as clean in Table \ref{tab:KS_IC_ER} and \ref{tab:ASR}), we also tested the models under our distorted setting (2-dist). Furthermore, to evaluate the robustness of models to unseen distortions, the two noise datasets, FSD50k and DNS, are held out from the training phase among all the experiments and are only adopted during testing to create a domain mismatch scenario. Note that speech in the FSD50k-distorted testing set (denoted as fsd in Table \ref{tab:KS_IC_ER} and \ref{tab:ASR}) contains one background noise sampled from the FSD50k corpus, creating a single distortion setting. Speech in the DNS-distorted testing set (denoted as dns in Table \ref{tab:KS_IC_ER} and \ref{tab:ASR}) is constructed by adding one background noise and convolving a room impulse response \cite{allen1979image} to speech.

For non-additive distortions, we chose some common sound effects, such as reverberation, pitch shift, and band rejection to apply to speech data. Adding non-additive distortions to speech does not require additional data and can be directly applied to waveforms. We followed the configurations and implementation details proposed in WavAugment\footnote{\href{https://github.com/facebookresearch/WavAugment}{https://github.com/facebookresearch/WavAugment}} \cite{wavaugment2020}. 

For the \textbf{ASR} task, we also report the performance on the test-other split of LibriSpeech and the real speech recordings of CHiME3 \cite{barker2015third}, no additional distortions are applied to these two testing sets.

\subsection{Upstream models and training details}
HuBERT-base is the teacher model used for knowledge distillation, and the pre-trained weights are initialized by the checkpoints released in Fairseq\footnote{\href{https://github.com/facebookresearch/fairseq}{https://github.com/facebookresearch/fairseq}} \cite{ott2019fairseq}. The domain-adaptive pre-trained version of HuBERT-base follows the same procedure mentioned in \cite{huang2022improving}, except for the generating process of distorted speech replaced by our procedure mentioned in Section \ref{ssec:dataprep}. 

For knowledge distillation, we train each model for 20k steps and adopt the checkpoint that yields the lowest distillation loss (Eq.(\ref{eq:distil_loss})) on the development set (dev-clean of LibriSpeech). Other hyperparameters such as learning rate, optimizers, and schedulers are the same as the original DistilHuBERT \cite{chang2022distilhubert} training configuration.

For DAT, the distortion classifier is a mean pooling operation followed by a linear layer projecting the representations to a dimension equal to the number of distortion types. In our work, there are seven distortion types, including Musan, Gaussian, WHAM!, reverberation, pitch shift, band rejection, and clean. The distortion classifier is trained in a multi-label classification style. The $\lambda$ value is set to $1\mathrm{e}{-2}$ for all of the DAT experiments.

\subsection{Downstream models and training details}
For the downstream speech processing models of the four tasks reported in the results, we follow the same model configurations of the SUPERB benchmark\footnote{Details for training downstream speech models can be found at \href{https://github.com/s3prl/s3prl/blob/main/s3prl/downstream/docs/superb.md}{https://github.com/s3prl/s3prl/blob/main/s3prl/downstream/docs/superb.md}}. We adopt the last hidden states of the student model as the input of the downstream models. During downstream training, the batch sizes for training downstream tasks \textbf{KS}, \textbf{IC}, and \textbf{ASR} are
set to 32, and 4 for \textbf{ER}. The learning rate for the optimizer is set to $1\mathrm{e}{-4}$ for \textbf{IC}, \textbf{ER}, and \textbf{ASR}, and $1\mathrm{e}{-3}$ for \textbf{KS}. Tasks \textbf{KS}, \textbf{IC}, and \textbf{ASR} are trained for 200k
steps and task \textbf{ER} is trained for 30k steps.

\begin{figure*}[ht]
\centering  
  \begin{subfigure}[b]{.19\linewidth}
    \centering
    \includegraphics[width=.9\textwidth]{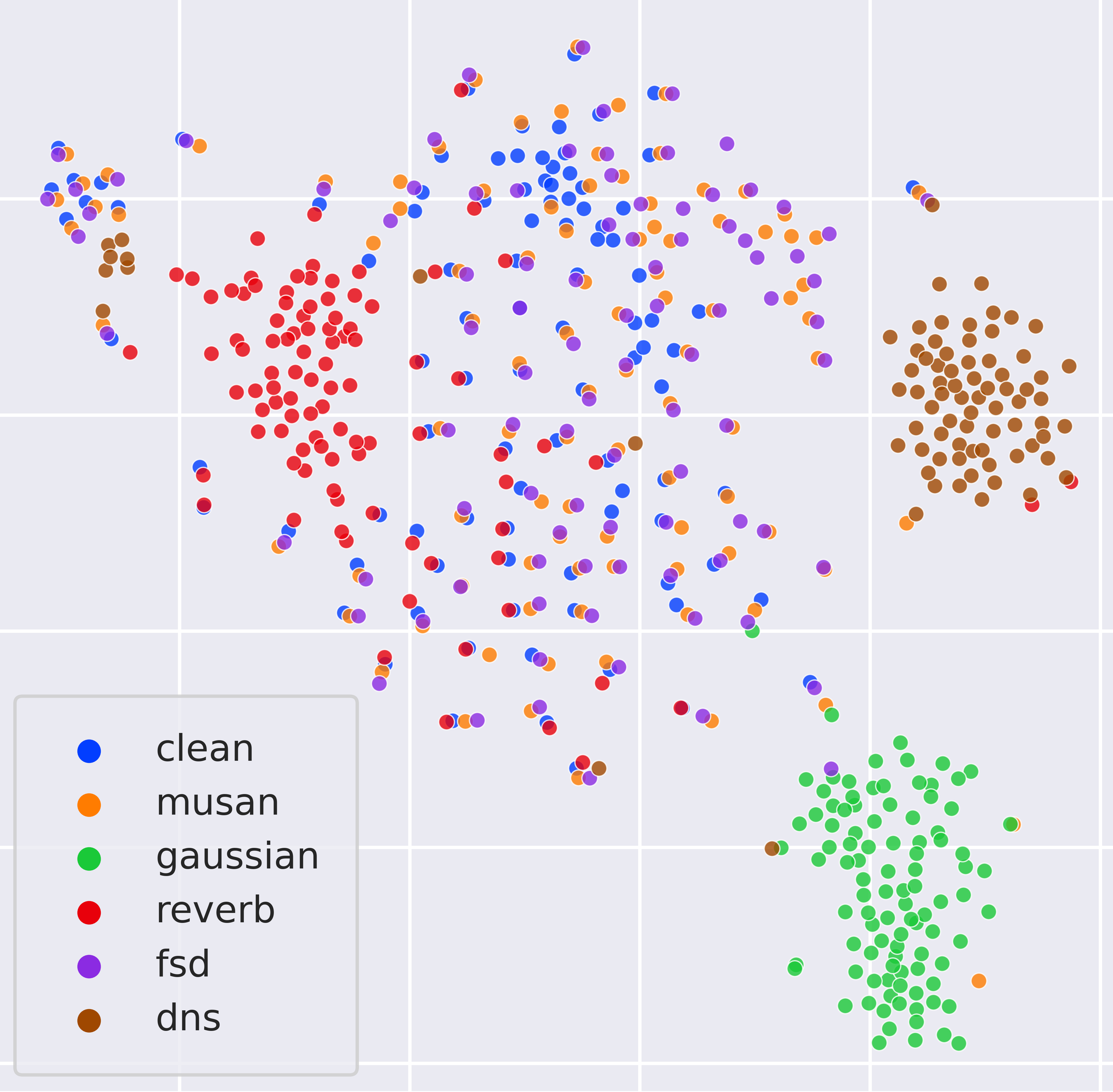}
    \caption{(T1)}\label{fig:T1}
  \end{subfigure}
  \hfill
  \begin{subfigure}[b]{.19\linewidth}
    \centering
    \includegraphics[width=.9\textwidth]{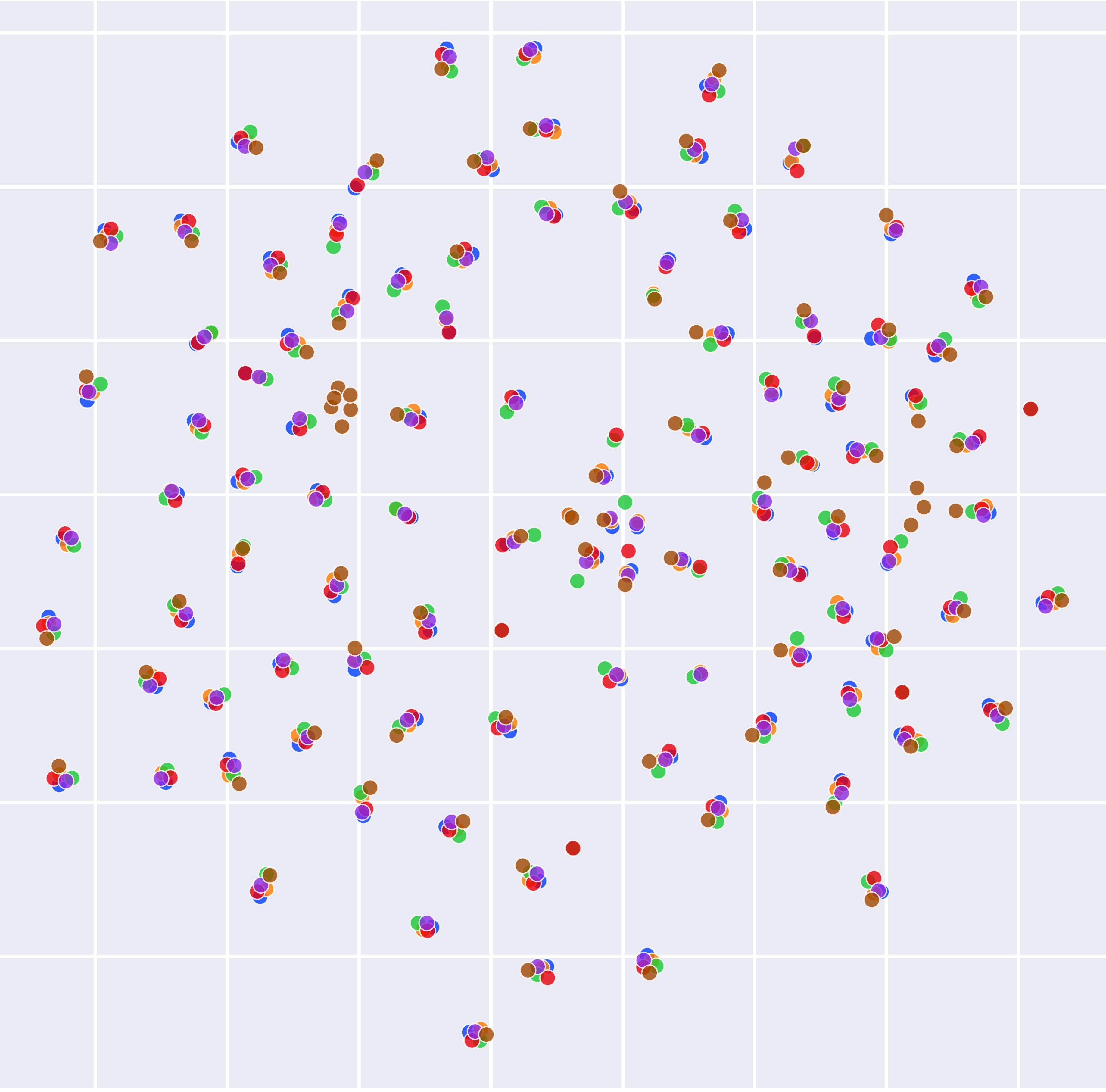}
    \caption{(T1')}\label{fig:T1'}
  \end{subfigure}
  \hfill
    \begin{subfigure}[b]{.19\linewidth}
    \centering
    \includegraphics[width=.9\textwidth]{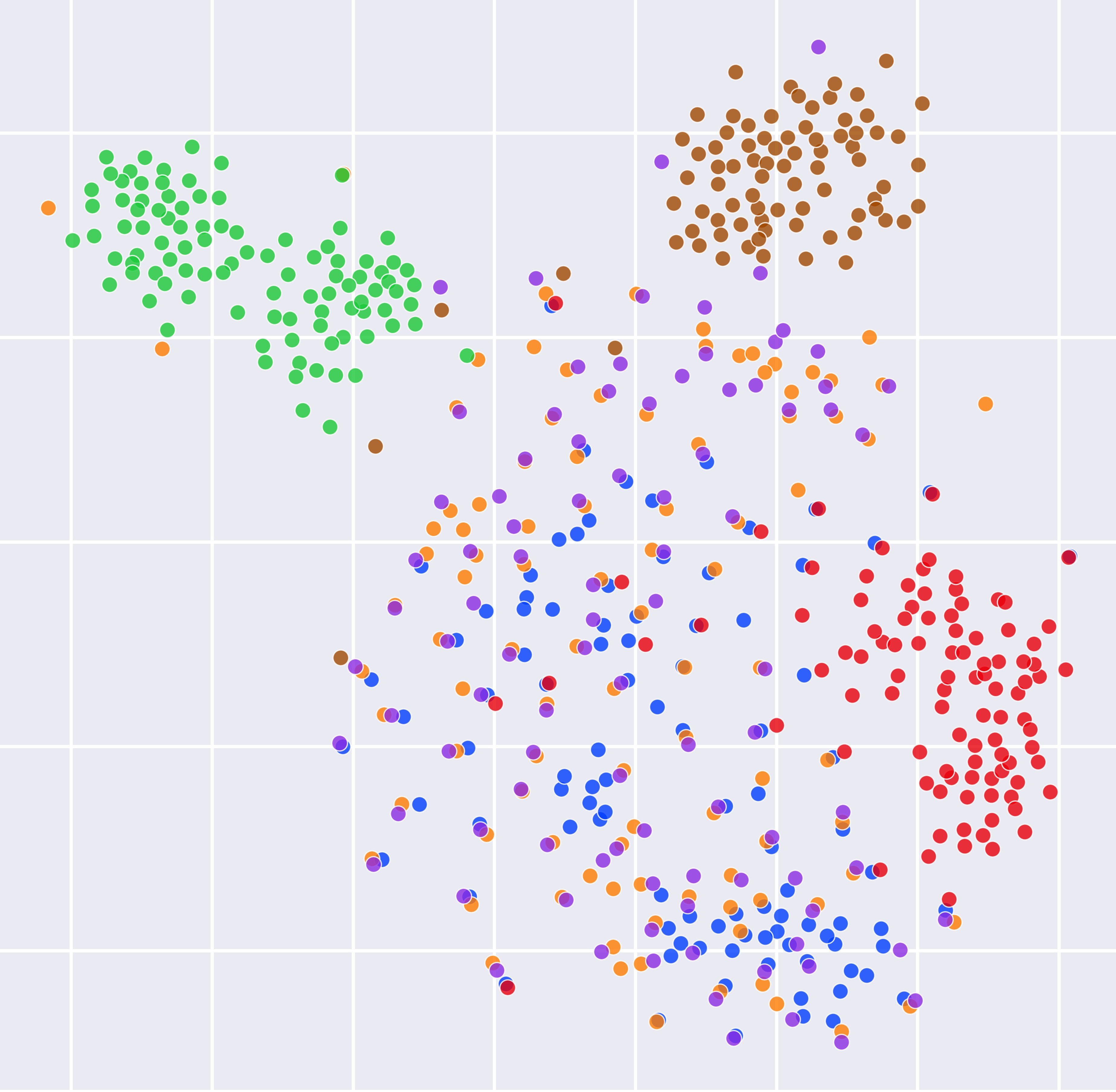}
    \caption{(S1)}\label{fig:S1}
  \end{subfigure}
  \hfill
  \begin{subfigure}[b]{.19\linewidth}
    \centering
    \includegraphics[width=.9\textwidth]{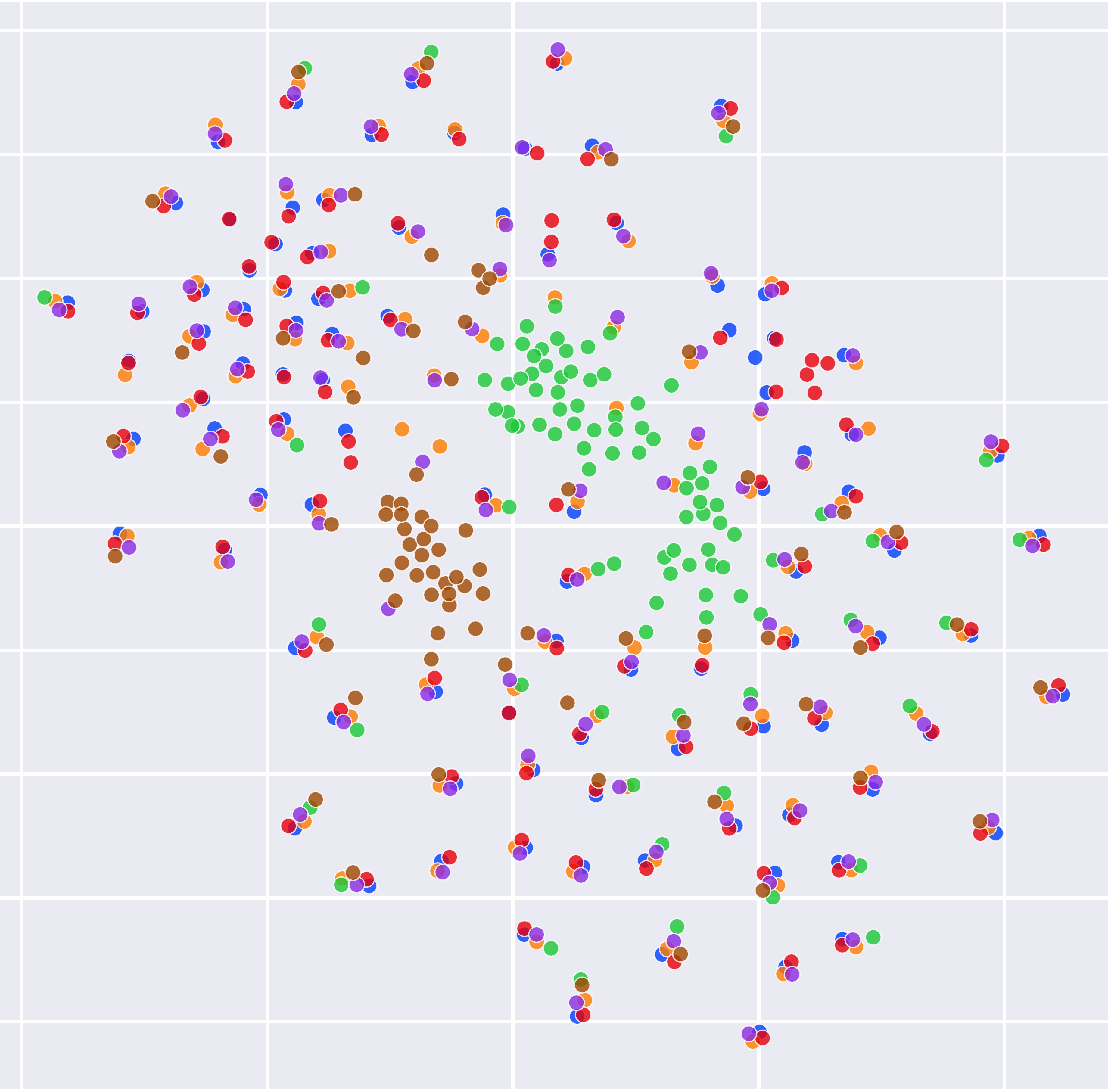}
    \caption{(S1')}\label{fig:S1'}
  \end{subfigure}
    \hfill
  \begin{subfigure}[b]{.19\linewidth}
    \centering
    \includegraphics[width=.9\textwidth]{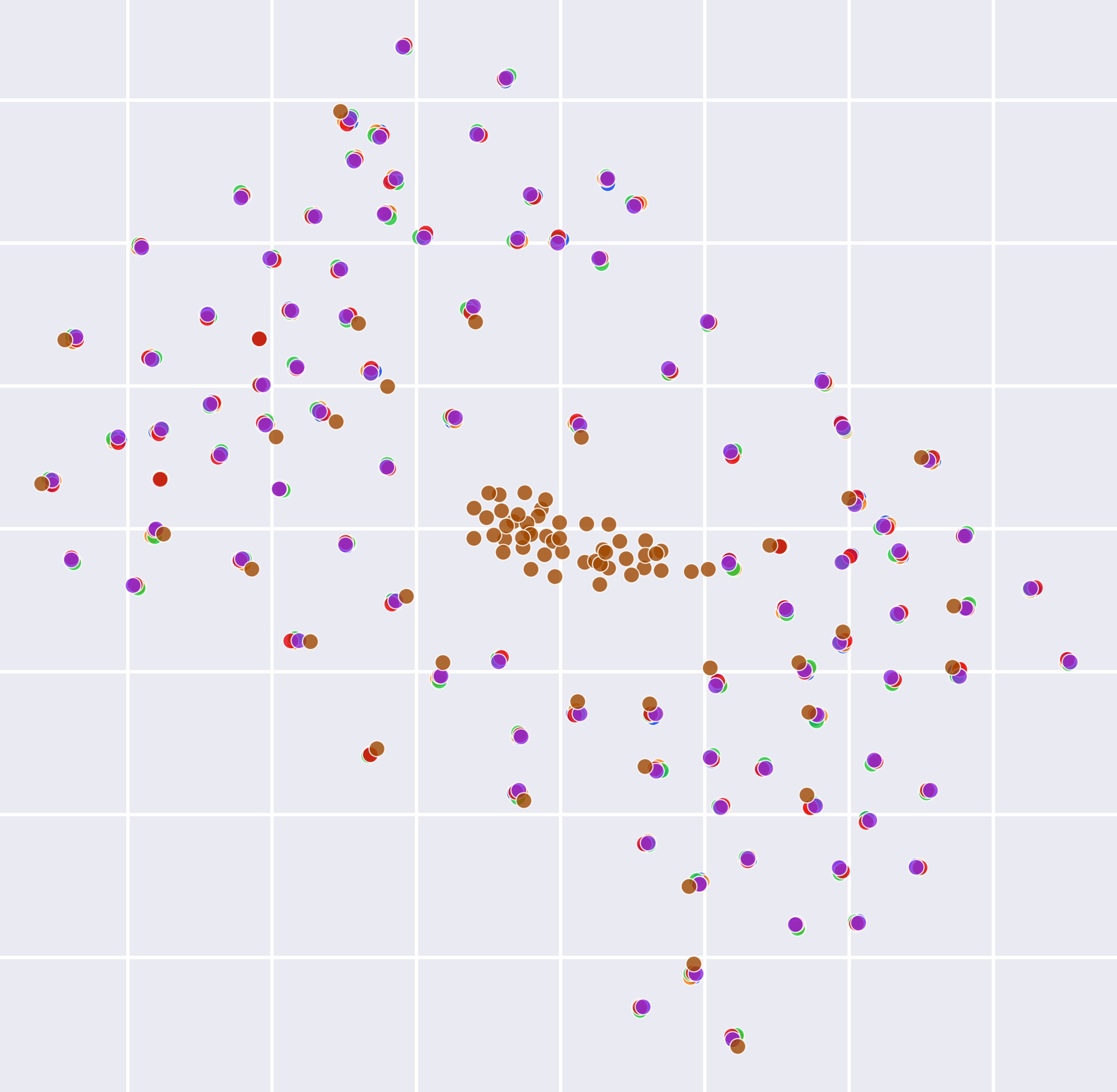}
    \caption{(S2)}\label{fig:S2}
  \end{subfigure}%
  \hfill
    \begin{subfigure}[b]{.19\linewidth}
    \centering
    \includegraphics[width=.9\textwidth]{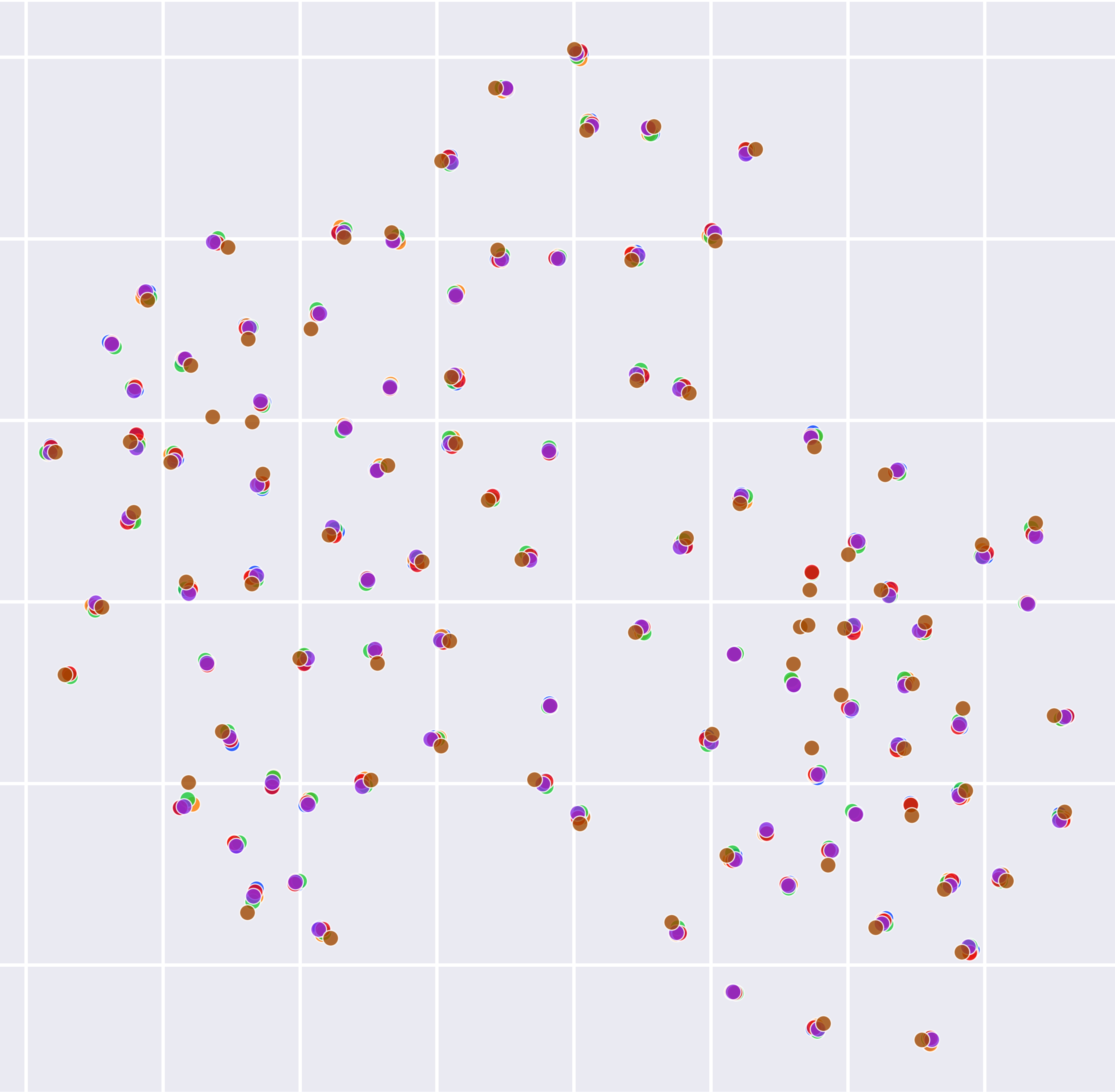}
    \caption{(S2')}\label{fig:S2'}
  \end{subfigure}%
  \hfill
    \begin{subfigure}[b]{.19\linewidth}
    \centering
    \includegraphics[width=.9\textwidth]{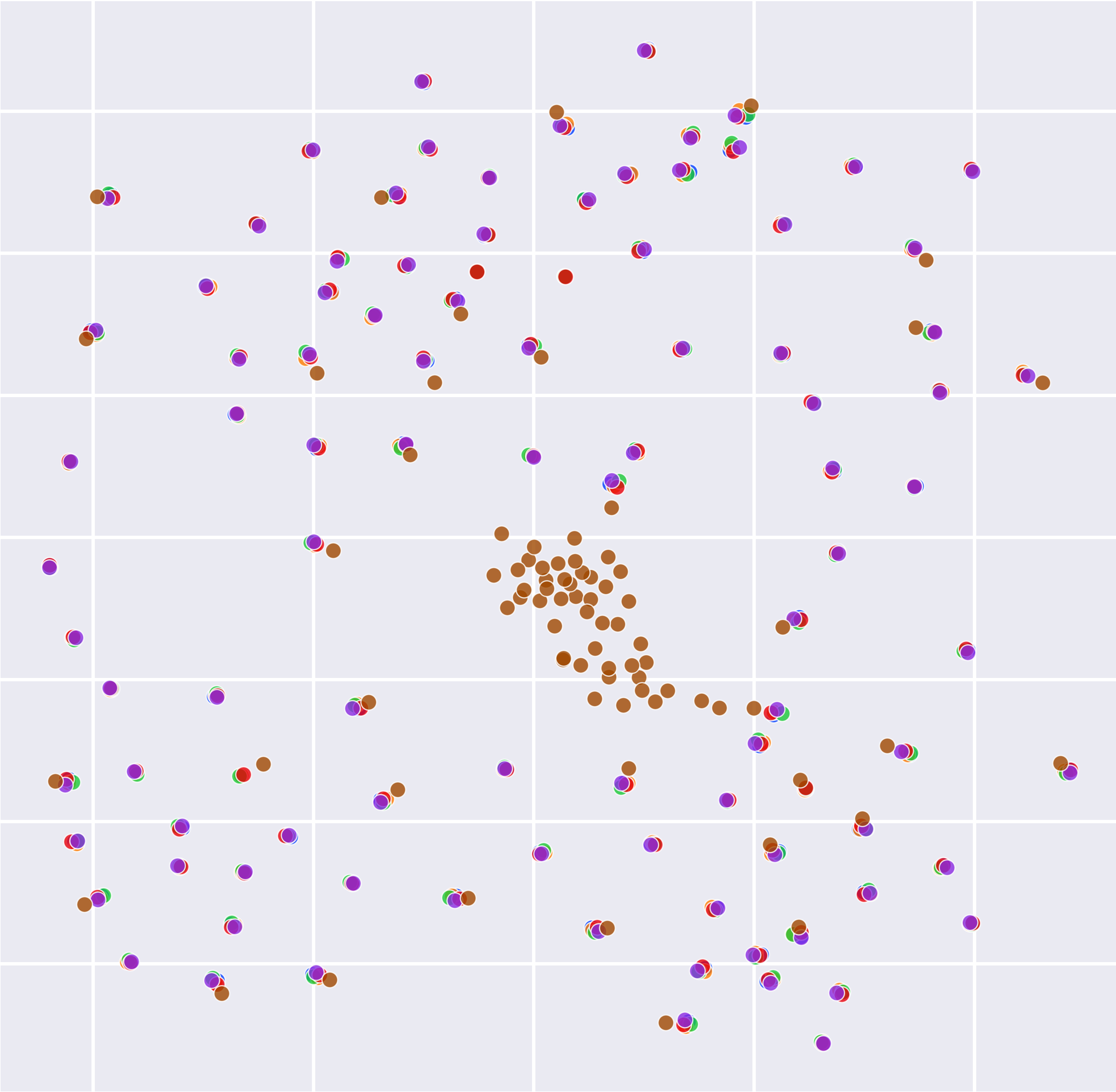}
    \caption{(S4)}\label{fig:S4}
  \end{subfigure}%
    \hfill
  \begin{subfigure}[b]{.19\linewidth}
    \centering
    \includegraphics[width=.9\textwidth]{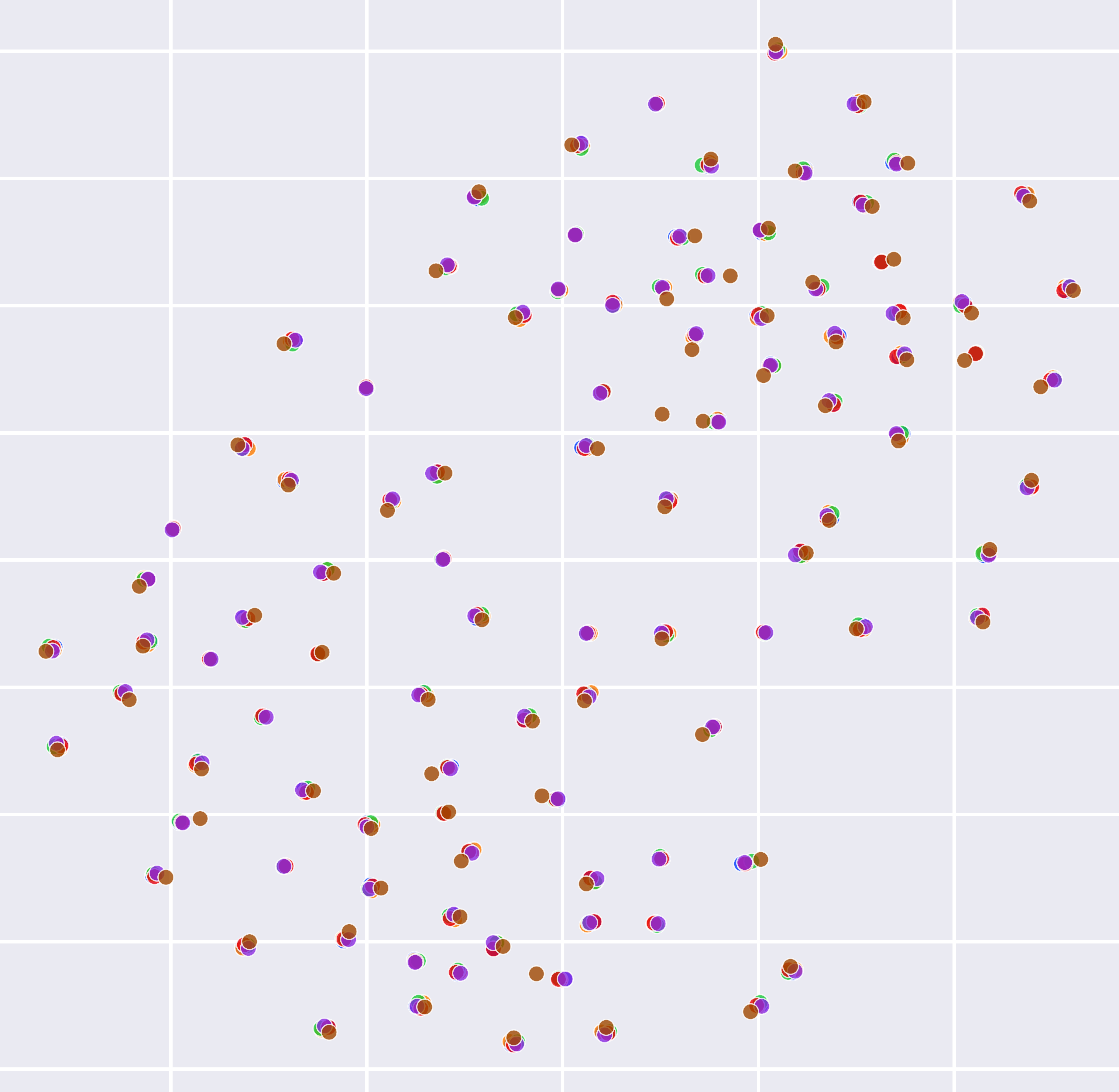}
    \caption{(S4')}\label{fig:S4'}
  \end{subfigure}%
  \hfill
    \begin{subfigure}[b]{.19\linewidth}
    \centering
    \includegraphics[width=.9\textwidth]{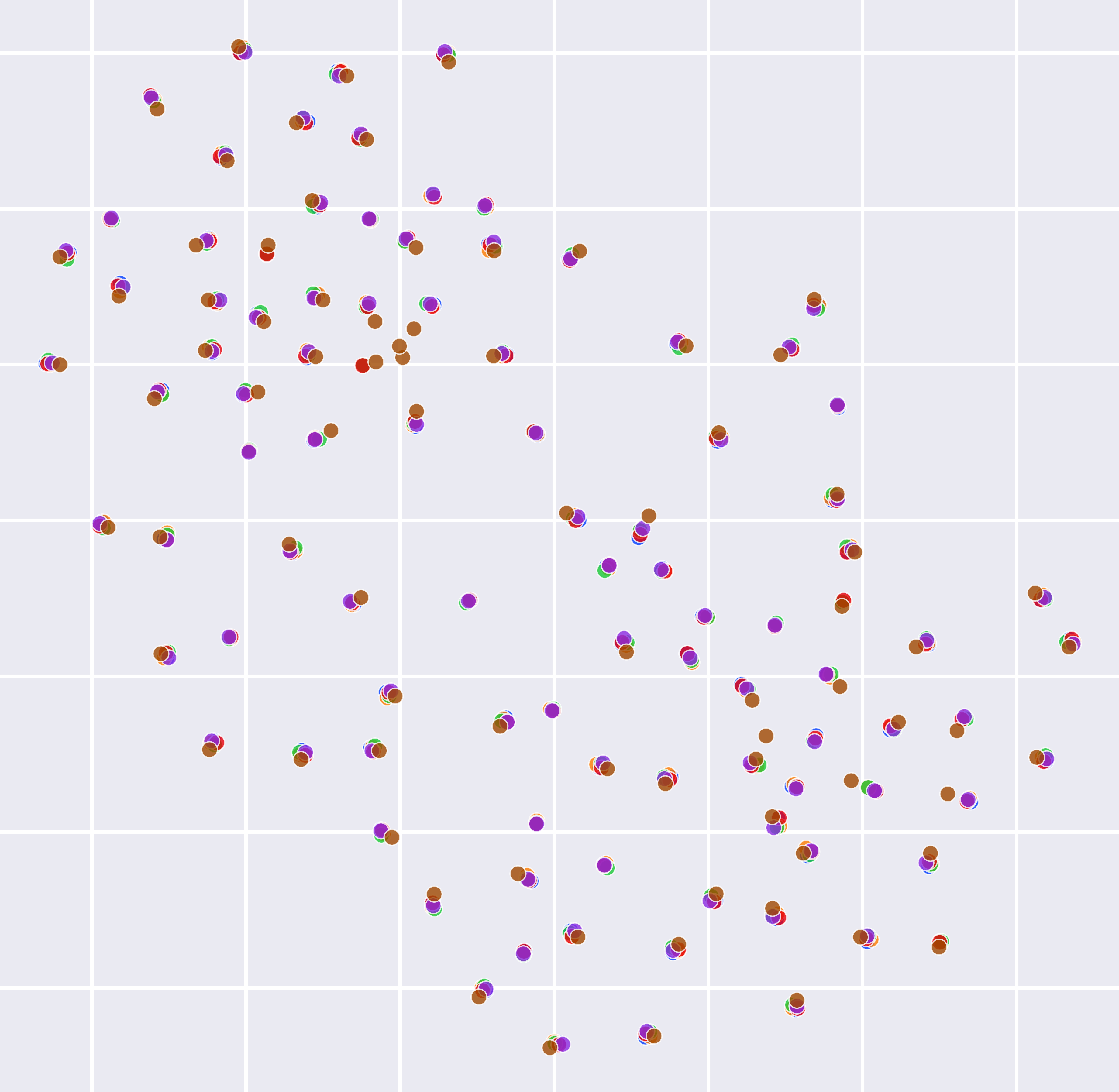}
    \caption{(S5)}\label{fig:S5}
  \end{subfigure}%
  \hfill
    \begin{subfigure}[b]{.19\linewidth}
    \centering
    \includegraphics[width=.9\textwidth]{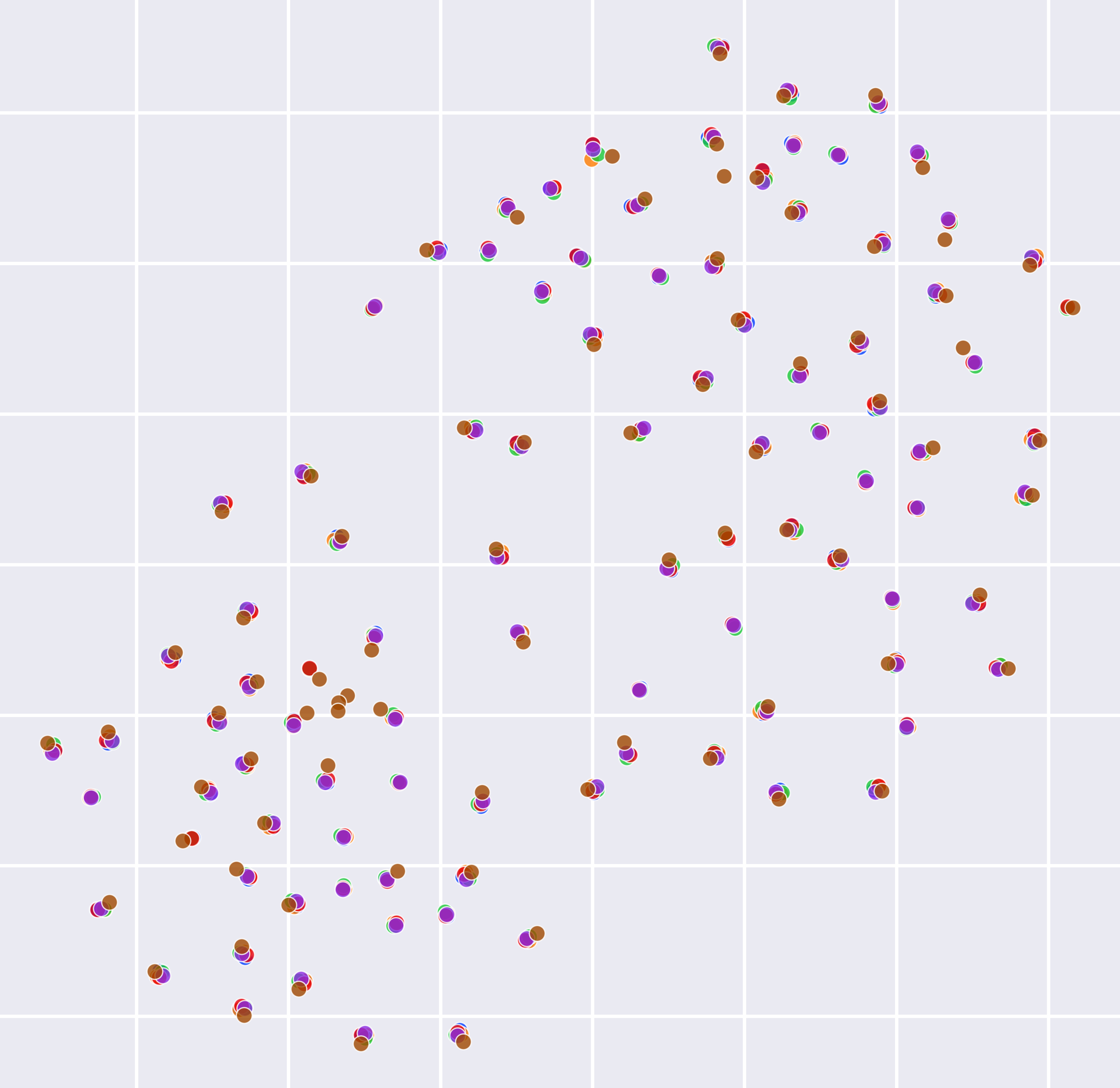}
    \caption{(S5')}\label{fig:S5'}
  \end{subfigure}%
  \caption{Visualization of representations for models in Tables \ref{tab:KS_IC_ER} and \ref{tab:ASR}. Each visualization corresponds to an upstream model without any fine-tuning with downstream data. Colors blue, orange, green, red, purple, and brown represent the representations of clean speech and speech with Musan, Gaussian, reverberation, FSD50k and DNS noises, respectively. FSD50k and DNS noises are unseen distortions during pre-training.}
\label{fig:embeddings}
\end{figure*}

\section{Results}
\label{sec:results}

\subsection{Baselines: HuBERT and DistilHuBERT}
By comparing (T1) to (S1), and (T1') to (S1') in Table \ref{tab:KS_IC_ER} and Table \ref{tab:ASR}, it is obvious that HuBERT and DistilHuBERT have similar performance on the clean testing set for most of the speech tasks. However, both models suffer from performance degradation when distortions are introduced, especially for DistilHuBERT. This suggests that HuBERT is not robust, and the distillation process even worsens the generalizability of it.


\subsection{Different CDM settings}
Setup1 forces the student to learn the clean representations of the teacher regardless of the distortions of speech. The results show that applying setup1 by forcing representations of all kinds of input speech (clean or distorted) to fall in the clean representation domain is effective ((S2)(S2') compared to (S1)(S1')). However, by comparing (S2) to (S4), and (S2') to (S4'), we observe that setup2 yields better performance than setup1 on almost every testing set for \textbf{KS}, \textbf{IC}, and \textbf{ER}. This indicates that learning representations of the same speech utterance but with different distortions improves the generalizability of the student model. 

For setup2, we also experimented on the case where the teacher and student models take the same distorted speech as input ((S3) and (S3')) and found out that this setting yields low distillation loss but does not generalize as well as the original setup2 on each testing set. We also observed that (S3) and (S4) have large performance gaps on the FSD50k and DNS testing set for \textbf{IC}, \textbf{ER}, and \textbf{ASR}. This is because the student model (S3) is prone to output representations containing distorted information when the teacher and student have the same distorted speech inputs, causing the representations to be less domain-invariant. 

\subsection{Domain Adversarial Training}
\label{subsec:dat_results}
Setup1 with DAT not only forces the student to map its representations to the clean representations of the teacher, but also regulates the representations of the last hidden layer to be domain-invariant. The results show that regulating the representations of the last hidden layer is effective for both distorted and clean speech ((S5) (S5') compared to (S2) (S2')), and this setting (S5) even outperforms setup2 (S4) on most of the testing sets. We notice that whether using the domain-adaptive pre-trained HuBERT as the teacher model seems to have minor impacts on the average performances by comparing (S5) and (S5'), implying that DAT reduces the gap between different teacher models (the gap between (S5) and (S5') compared to the gap between (S2) and (S2')).

We constructed models (S6) and (S6') by applying DAT to models (S4) and (S4'). By comparing (S6) with (S4), it showed that applying DAT to setup2 improved the performance for \textbf{IC}. By comparing (S6') with (S4'), we also found similar results for \textbf{ER} and \textbf{ASR}. However, we noticed that there were still some cases where models did not benefit from DAT by observing (S4) and (S6) having similar results for \textbf{KS}, \textbf{ER}, and \textbf{ASR}. Knowing that model (S4) does not use the domain-adaptive pre-trained HuBERT model during distillation, causing a performance gap between (S4) and (S4'), we hoped that applying DAT to (S4) could make up the gap. Unfortunately, this is not the case. DAT also seems to worsen the performances of some testing sets of \textbf{KS} and \textbf{IC}. 


\subsection{Different model sizes}
We also trained different model configurations of DistilHuBERT by alternating the number of the transformer encoder layers of the student model. 
From models (S7)(S7')(S8)(S8'), by comparing the performance between clean testing sets and distorted testing sets (2-dist, fsd, and dns), we found out that smaller models are less robust to distortions.
To ensure general usage of our proposed methods, we trained a smaller student model (S7') and a larger student model (S8') under the setting that yielded best performance (setup2). 
By comparing (S7) to (S7'), and (S8) to (S8'), we conclude that setup2 shows consistent results for student models of different sizes. This demonstrates that this setting is model-agnostic, and can be applied to different student architectures in the future. 

\section{Visualization}

\subsection{Visualization setup}
\label{ssec:visual}
To demonstrate the robustness of our proposed approaches, we visualized the last layer representations of the models with t-SNE \cite{vanDerMaaten2008} for the test-clean portion of LibriSpeech. We show the speech representations of six kinds of speech, including clean speech, speech with Musan noise, speech with Gaussian noise, speech with reverberation, and speech distorted with FSD50k and DNS noise by the following process. First, we distort all the speech utterances in the test-clean set with one kind of speech distortion and extract their representations from the last transformer layer of the upstream model. The representations are further averaged along the timestep dimension to produce a flat vector of length $D$. Then we divide the representations into 100 splits and average the representations in each split, resulting in 100 representations. Finally, we repeat this process for the six kinds of speech, resulting in 6 vectors for each split and 600 vectors in total for visualization.

\subsection{Visualizing results}
\label{ssec:visual_results}
We show t-SNE visualizations as described in Section~\ref{ssec:visual} to further understand the modeling capability of each upstream model. From Fig.~\ref{fig:T1}, we observe that the original HuBERT model (T1) is not robust when the speech signal is subdued under different distortions. There are clear cluster assignments for each of the distorted speech configurations. The two most prominent clusters are the ones with Gaussian noise and DNS noise added to speech. This explains the large performance gap of model (T1) between the clean testing set and the testing set with DNS noise in Tables \ref{tab:KS_IC_ER} and \ref{tab:ASR}.  A similar phenomenon can be seen for the model distilled from the original HuBERT model (S1) (see Fig.~\ref{fig:S1}). 

Fig.~\ref{fig:T1'} shows that model (T1') is robust to all the distortions. Notice that some figures in Fig.~\ref{fig:embeddings} show multiple data points overlapped together. We verified that the representations representing speech with clean, Musan, Gaussian, reverberation, FSD50k, or DNS noise in the same split overlap together on the t-SNE visualization, meaning that, no matter if the speech signals belonging to a particular split have been distorted or not, the model will still place them into the same representational space. This supports our claim that model (T1') is more robust than the baseline teacher model (T1).
Having a robust teacher model proves to be crucial, as can be seen by comparing Fig.~\ref{fig:S1} and \ref{fig:S1'}, where we show that a model distilled from a teacher that is not robust results in less robust student models and vice versa, explaining the large performance gap between models (S1) and (S1') in Table \ref{tab:KS_IC_ER} and Table \ref{tab:ASR}. 
Similar conclusions can be inferred for models in Fig. \ref{fig:S2} and \ref{fig:S2'}. 

Finally, in Fig.~\ref{fig:S4'}, all the points in the same split with different distortions almost completely overlap with each other, showing the distortion-invariant capability of performing distillation with CDM. On the other hand, Fig. \ref{fig:S5} and \ref{fig:S5'} show that DAT is important when the teacher model is not robust, as supported by the analysis in section \ref{subsec:dat_results} and the results for models (S5) and (S5') in Table \ref{tab:KS_IC_ER} and Table \ref{tab:ASR}. Yet, performing DAT does not always improve the robustness of the student when distilled from a robust teacher model. Hence, performing distillation with our CDM method is enough for achieving robustness under distorted settings when we have a robust teacher. These conclusions are supported by the results in Table \ref{tab:KS_IC_ER} and Table \ref{tab:ASR} where the difference in performance between model (S4') and (S6') are almost marginal. Our visualization provides insightful understandings of which pipeline to use depending on the characteristics of the teacher model involved during the distillation process.


\section{Conclusion}

In this paper, we find that though having similar performance as original SSL models, distilled SSL models suffer from performance degradation even more than their original versions in distorted environments. 
This paper proposes to apply two different setups of CDM during distillation to improve the generalizability of distilled SSL models. We found that models trained under the second setup, which applies different distortions to the speech input for the teacher and student model, tend to yield more distortion-invariant representations than models trained under the first setup. We also show that depending on the characteristics of the teacher model, using only CDM during distillation is enough, yet adding DAT improves student generalization performance when the teacher model is not robust.
Results show consistent improvements under both in- and out-of-domain distorted setups for different downstream tasks while keeping efficient model size. 

\section{ACKNOWLEDGMENTS}
\label{sec:ack}

Part of the work presented here was carried out during the 2022 Jelinek Memorial Summer Workshop on Speech and Language Technologies at Johns Hopkins University, which was supported with unrestricted gifts from Amazon, Microsoft, and Google. We also thank to National Center for High-performance Computing (NCHC) and Taiwan Web Service (TWS) for providing computational and storage resources.

\bibliographystyle{IEEEbib}
\bibliography{strings,refs}

\end{document}